\documentclass[journal]{IEEEtran/IEEEtran}

\ifCLASSINFOpdf
\else
\fi
\usepackage{dblfloatfix}
\usepackage{graphicx}
\usepackage{color}
\usepackage[x11names]{xcolor} 
\usepackage{soul}
\usepackage{verbatim} 

\usepackage{textcomp}

\usepackage{framed}

\usepackage{fixltx2e} 

\usepackage{subfigure}
\usepackage{amsmath} 
\usepackage{url}

\usepackage[normalem]{ulem}

\newcommand{\AP}[1]{{\color{blue}{[}
{AP: #1}{]}}}
\newcommand{\SR}[1]{{\color{olive}{[}
{SR: #1}{]}}}
\newcommand{\MH}[1]{{\color{red}{[}
{MH: #1}{]}}}
\newcommand{\SK}[1]{{\color{violet}{[}
{SHK: #1}{]}}}
\newcommand{\SG}[1]{{\color{teal}{[}
{SG: #1}{]}}}

\begin{document}
%
\title{
Dark Memory and  Accelerator-Rich System \\
Optimization in the Dark Silicon Era
}
%
%
%

\author{Ardavan Pedram,~\IEEEmembership{Member,~IEEE,}
        Stephen Richardson,~\IEEEmembership{Member,~IEEE,}\\
        Sameh Galal,~\IEEEmembership{Member,~IEEE,}
        Shahar Kvatinsky,~\IEEEmembership{Member,~IEEE,}
        and~Mark Horowitz,~\IEEEmembership{Fellow,~IEEE}
\thanks{Ardavan Pedram, Stephen Richardson and Mark Horowitz are with the 
Department of Electrical Engineering, 
Stanford University, Stanford,
California.}
\thanks{Sameh Galal is with Soft Machines Inc, Santa Clara, California.}
\thanks{Shahar Kvatinsky is with the Department of Electrical Engineering, Technion - Israel Institute of Technology, Haifa, Israel.}
\thanks{Manuscript received February 2016; revised April 2016.}}

%
%

\markboth{IEEE Design \& Test,~Vol.~XX, No.~XX, XXXXX~2016}%
{Shell \MakeLowercase{\textit{et al.}}: Bare Demo of IEEEtran.cls for IEEE Journals}
%



\maketitle


    \begin{abstract}


The key challenge to improving performance in the age of Dark Silicon is how
to leverage transistors when they cannot all be used at the same time.  In modern 
system-on-chip (SoC) devices, 
these transistors are often 
used to create specialized accelerators which  improve energy efficiency for some applications by 10-1000X. While this might seem like the magic bullet we need, for most CPU applications more energy is dissipated in the memory system than in the processor: these large gains in efficiency are only possible if the DRAM and memory hierarchy are mostly idle. We refer to this desirable state as Dark Memory, and it only occurs for applications with an extreme form of locality.

To show our findings, we introduce Pareto curves in the energy/op and mm$^2$/(ops/s) metric space for compute units, accelerators, and on-chip memory/interconnect. These Pareto curves allow us to solve the power, performance, area constrained optimization problem to determine which accelerators should be used, and how to set their design parameters to optimize the system. It also shows that memory accesses create a floor to the achievable energy-per-op. 
Thus high performance requires Dark Memory, which in turn requires co-design of the algorithm, for parallelism and locality, along with the hardware.

\end{abstract}


%
\IEEEpeerreviewmaketitle






\section{Introduction}

Even though Dennard in 1974 showed how to scale CMOS devices for constant power density 
as the feature size scaled down by a factor 
\textit{$\alpha$ = (newSize/prevSize)}~\cite{Dennard:jssc74},
the power density of
CMOS processor chips grew exponentially from the mid 80s to the late 90s. This power growth resulted
both from scaling clock frequency faster than 1/$\alpha$ 
and
voltages slower than $\alpha$~\cite{danowitz2012cpu}.  By the mid 2000’s this growing power meant that
all computing systems, even high-end servers, had become power limited. 
 
Unfortunately, during this period, voltage scaling essentially stopped.  Now, when moving to a technology 
with feature size scaled by $\alpha$ with respect to the previous generation,
gate energy scales by at best $\alpha$ 
(not $\alpha^3$ as before).
So even when we do not scale clock frequency at all and just try to build 
$\alpha^{\textrm{--}2}$ 
processors (to use all transistors available in the same area), the power will increase by 
$\alpha^{\textrm{--}1}$ 
which will exceed the power budget. This inability to use, or at least use concurrently,  all the gates you can create on a silicon die gave rise to the term Dark Silicon~\cite{esmaeilzadeh2011dark}.



Today the key challenge in improving performance is how to leverage transistors when they cannot all be used at the same time. Michael Taylor, in his \textit{Four Horsemen of Dark Silicon} paper, characterized the work in this field into four different approaches: shrink, dim, specialize, and technology magic~\cite{taylor2012dark}.

The simplest approach is 
to simply not build transistors that cannot be used continuously:
only build the number
of gates that you can operate concurrently
under a given power constraint.
Since this number is growing slowly, the resulting die area will shrink with technology
scaling. This is the \textit{shrink} horseman. While the power density 
of the silicon die does go up as area shrinks,
%
getting power out of the die is not the main problem, e.g. heat pipes work well for this.  The main problem is getting the power out of the complete system, whose form factor does not change when the die shrinks.
%
%
The shrink approach makes the computing device cheaper to manufacture, but significantly limits the performance
improvement.

\textit{Dim}, tries to leverage all the possible gates/transistors by making
some or all of them
dissipate less power than before. This dimming generally reduces the performance
per unit area, so it must be done in a way that results in better overall performance
than simple die shrinking.  
Two common dimming techniques are
lowering the supply voltage to reduce gate energy, and 
increasing the numbers of gates in a clock cycle to decrease the clock energy and the number of gate 
evaluations/sec.\footnote{Lowering the clock speed 
decreases the number of gate evaluations per second, 
but of course also lowers the performance.  The performance loss is often less than the change in clock frequency since the shorter pipeline generally has higher architectural efficiency and thus better energy efficiency.
}
Dimming techniques have been widely used to create today's multi-core processors,
and have grown quite sophisticated. For example many processors dynamically
adapt their supply voltage 
so aggressively
that they have to lower their clock frequency when they detect small power supply glitches~\cite{grenat20145}. We show
in Section~\ref{sec:metrics} that these techniques create Pareto curves in the
energy efficiency and compute density metric space. These curves together with
the design power, performance, and area constraints can be used
to determine the optimal amount of dimming. 

The next horseman, \textit{specialization}, 
uses the extra transistors to create compute engines highly optimized to specific
applications. This specialization can dramatically improve energy efficiency which, 
in a power limited world, enables higher performance. Since they run only specific
applications, these engines are idle, or dark, most of the time, a perfect fit to Dark Silicon constraints.
Specialized accelerators are widely used in modern processor 
systems-on-chip (SoCs) 
and many of these are orders of magnitude more energy efficient than a CPU or GPU. This
dramatic improvement in energy efficiency has led many people think this approach is the
key to designing a Dark Silicon chip. 

Yet when you look at power dissipation in a CPU chip, around half
is in the on-chip 
memory system~\cite{horowitz2014keynote}.
Re\-mem\-ber\-ing that most power limitations
are really system and not chip-level power limitations, this actually understates the memory problem, since
we should really include
external DRAM power as well. Thus the memory system contributes well over
50\% of the total system power.
So, given Amdahl's Law, changing the compute engine without improving the memory can only have a modest (less than 2x) change in energy efficiency. Section~\ref{sec:darkmemory} explores this issue in more detail, explaining why memory fetches are 
expensive, 
how their energy costs grow with memory size, and how to compute the lower bound on an application's energy consumption from the locality of the running application. The unavoidable conclusion is that high performance requires the DRAM and most of the lower levels of memory hierarchy~(e.g. last level cache) to be dark almost all of the time.  We 
call this idle memory \textit{Dark Memory.} Given this insight, Section~\ref{sec:codesign} describes the critical task for Dark Silicon systems: optimizing algorithms to maximize their exploitable locality.


Finally, the \textit{Deus Ex Machina} horseman 
deals with Dark Silicon by hoping for
a dramatic change in the underlying device technology.
While it would be great if a new and better technology/approach is found, 
we have at least two reasons not to count on it.
First, all new technologies take time to reach the manufacturing scale needed to affect computing; even if a new technology is created, it is a decade away from affecting volume computing devices.  Given that there is no serious competitor today, computing will use CMOS for at least another decade. Second, waiting for
a new ``magic" technology abdicates our role in 
helping to continuously improve
computing performance. So the rest of the paper focuses
%
on existing mainstream silicon computing,
%
though we will also look briefly at the effect of potential new technologies.

Section~\ref{sec:metrics}
ties everything together by describing two simple metrics, energy/op and mm$^2$/(op/s) which enable us to bring all these techniques into a single framework, and thus determine what amount of shrink, dim, and specialization is best for a given design, as well as  quantifying the importance of keeping the memory dark and finding 
optimal cache hierarchy sizes
for a given workload. One can use this framework to trade-off memory and specialized processors, as well as comparing two applications with different compute and locality patterns.

\section{Why Dark Memory is Essential}
\label{sec:darkmemory}

  \begin{table}
  \newcommand\mystrut{\rule{0pt}{7.5pt}}
   \centering
   
  \begin{tabular}[c]{|c|cc|cc|} \hline
	 {{Operation}} & 
	 \multicolumn{2}{c|}{ 16 bit (int)}&
	 \multicolumn{2}{c|}{ 64 bit (dp)}

\mystrut \\ \cline{2-5}

    & {E/op}& vs. Add & {E/op}& vs. Add \mystrut \\ \hline       
Add           & 0.18   &  1.0x   &  ~5   &  1x    \mystrut \\ 
Multiply      & 0.62   &  3.4x  & 20   &  4x    \mystrut \\ \hline
16-word RF    & 0.12   &  0.7x  & 0.34  & 0.07x  \mystrut \\ 
64-word RF    & 0.23   &  1.3x  & 0.42  &  0.08x  \mystrut \\ \hline
4K-word SRAM  & 8      &  44x   & 26  &  5.2x  \mystrut \\ 
32K-word SRAM & 11     &  61x   & 47   &  9.4x  \mystrut \\ \hline
DRAM          & 640    & 3556x  & 2560 & 512x \mystrut \\ \hline

\end{tabular}
\vspace{0.5\baselineskip}
   \caption{%
   Energy per op, in pJ, for various ops in 45 nm.  The second column in each group shows energy multiple vs. a single add operation.
   } 
   \label{tab:Comparison}
\end{table}

\begin{figure}[tb] 
   \centering
   \vspace{-20pt}
   \includegraphics[
     width=0.75\columnwidth]{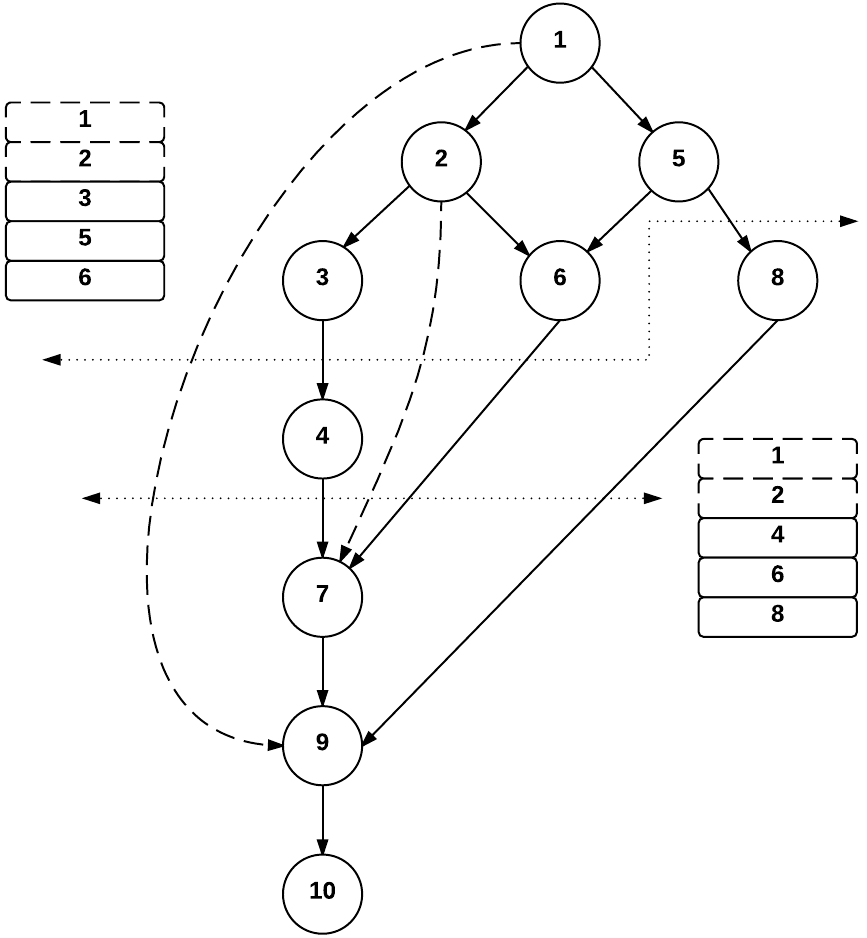} 
   \vspace{-10pt}
   \caption{Example data flow graph showing why wire length grows as communication become more complex. When this algorithm executes sequentially, the values on the wires that are still live in the execution trace are named and stored in a memory so they can be accessed when needed. The size of this live set sets the size of the required memory.  When new dependences are added from nodes 1 to 9 and nodes 2 to 7 (dashed lines), this adds two new registers 1 and 2 to the required memory at each indicated cut.}
   \label{fig:comm}
   \vspace{-1\baselineskip}
\end{figure}

The lesson that one quickly learns doing chip design today is that most of the energy is consumed
%
not in computation but in moving data to and from memory,
as can be seen in Table~\ref{tab:Comparison}.\footnote{Energy for memory and integer ops come from 
Verilog, placed and routed using commercial tools.
Energy for floating-point ops from Galal thesis~\cite{galal2012thesis} which also used data from placed and routed designs.} In ASIC style design, given e.g. the computational graph shown in Figure~\ref{fig:comm}, links between functional blocks are implemented as wires.  As the communication complexity grows, the hardware block gets bigger, making the wires longer, increasing the communication energy. 
Computers avoid this wire problem by serializing the computation, 
computing a few operations each cycle. However, it now needs to store the intermediate results, which used to
flow in wires 
from one logic unit to the next, so it can access them when needed.  Thus, the size of the memory needed
is related to the complexity of the communication in the algorithm, so the energy increases with communication complexity. 


\subsection{Memory Energy}

While the access energy of a memory 
depends on many factors, to first order it grows as the square root of its
size, which roughly corresponds to the length of the wires that need to transport the address and 
data values across the memory 
array.  
This was noted long ago by Amrutur and Horowitz~\cite{amrutur2000speed}, citing even earlier work by Evans and Franzon~\cite{evans1995energy}.
%
The memory energy also depends on the fetch width, but that dependence is much weaker than you might expect.  For example moving from 16- to 64-bit fetches only changes the energy by 1.5x, so wider fetches are generally more efficient in terms of energy per byte.
  \footnote{Internally
  most SRAMs fetch 64 to 256 bits on each access,
  so returning a small number of bits increases the effective energy cost per bit. To address this issue you could  create a SIMD machine and fetch the 16-bit data for four lanes from a single SRAM.  While this is more efficient, it also makes the memory 4x larger, since it now needs to hold four lanes' worth of working set, so the benefit is modest.
  }
This means that for a 16 bit machine, a fetch from even a 
4K-entry memory block costs over 10x the energy of a multiply operation,
as we saw in 
Table~\ref{tab:Comparison}.%

To minimize memory energy costs and improve performance, we create memory hierarchies, so that most of the accesses can be satisfied by small local memories.  Given that a normal operation requires 3 data fetches---two operands and one result---it is essential that the register file energy be as small as possible. The register energy is significant: for 16 bit arithmetic, the cost of these 3 fetches, 36pJ, exceeds the cost of a simple add operation, 18pJ, even when using a small 16 entry register file.%
\footnotemark[\value{footnote}]

The situation is actually worse than this since the actual cost of the operand fetch is higher than just the register file energy. It also took energy to load the value into, and store the result from, the register file.  This additional energy is set by the number of ops performed per register file load instruction, and grows as the register file gets smaller. 
This additional energy cost from needing to ``load/store" values from/to  a lower~(slower) level in the memory hierarchy exists until you get to DRAM, and can be significant:
since
the energy of a DRAM access is often two orders of magnitude larger than 
a local memory access,
the overall hit rate of the on-chip memory system needs to be better than 99\% for the DRAM not to dominate the overall memory energy.

While this seems to argue that larger memory hierarchies are better, 
both die cost and leakage constrain memory size. The problem is that while idle SRAM
may be dim, it is never completely dark. 
Each memory cell has a small leakage current
such that SRAM dissipates
static power, which can be a large issue for a battery operated
device. If the average activity of the device is low, minimizing this leakage moves
the optimal point to smaller memory sizes, 
which increases DRAM activity and results in a higher energy cost for each
memory access.\footnote{Another option is to power down the on-chip memory during idle
periods, but this too increases 
overall
memory energy since now the dirty cache
data needs to be written to DRAM on power-down, and additional DRAM fetches are needed to bring
the data back into the cache when it is powered back on.} 
%
Leakage energy and access energy both increase as the memory gets larger, and this leads to 
a minimum memory cost, which
is set by the application's locality. 
Section~\ref{sec:codesign}
shows methods to improve the locality of the algorithm you use and
Section~\ref{sec:metrics} shows how to find an optimal memory hierarchy for this improved algorithm.

\begin{figure}[tb] 
   \centering
   \vspace{-5pt}
   \includegraphics[
     width=\columnwidth]{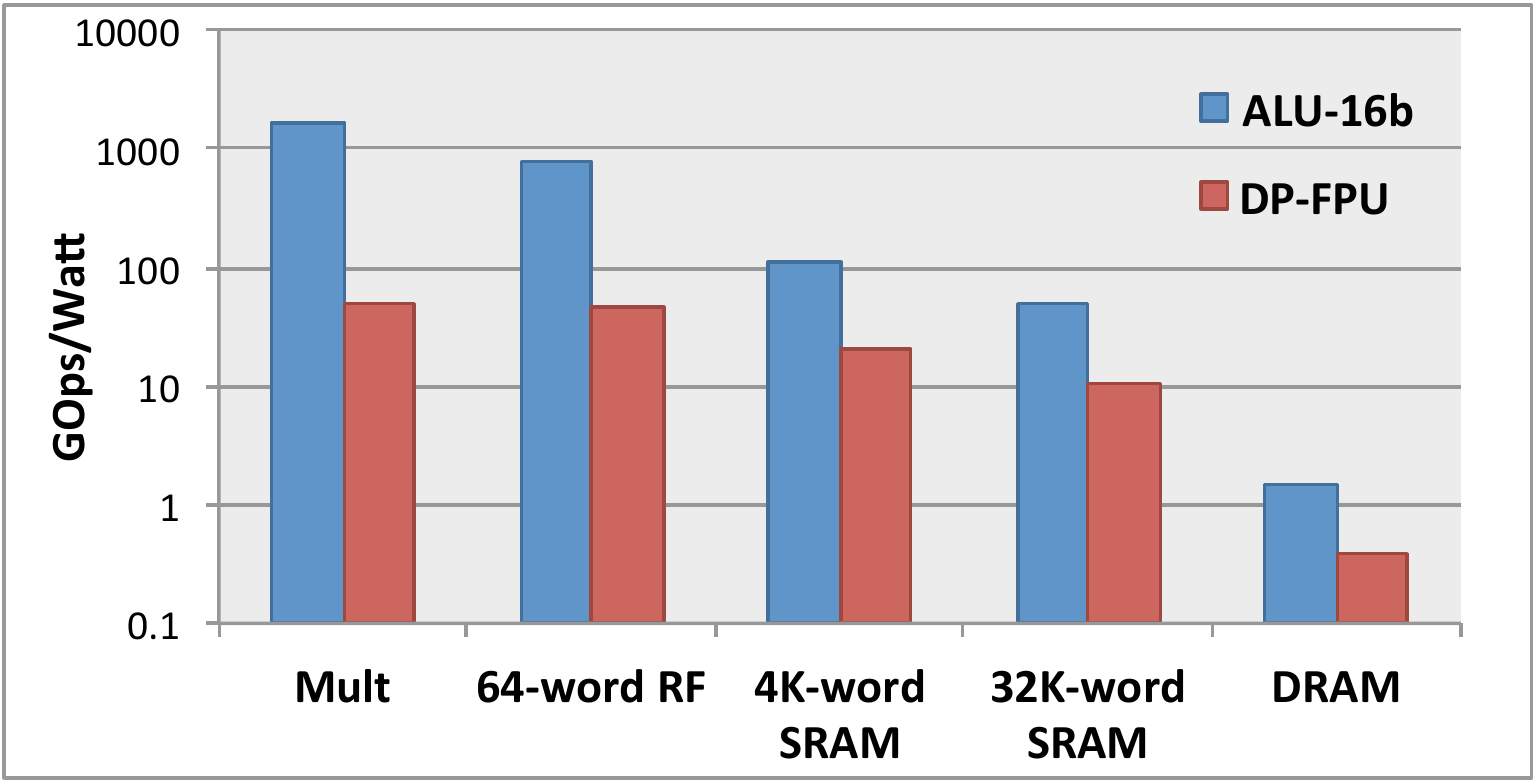} 
       \vspace{-20pt}
   \caption{Effective number of ops/s/W (ops/J) if one operand for that operation is fetched from the indicated memory, and the others come from the register file. For 16bit ops even a small 4Kword memory throttles the performance/W.}
   \label{fig:mem_pj}
   \vspace{-5pt}
\end{figure}

Another way to view 
memory's energy constraint
is shown in 
Figure~\ref{fig:mem_pj},
derived from the energy numbers of Table~\ref{tab:Comparison}.
Figure~\ref{fig:mem_pj} plots the maximum number of operations/sec for a watt of power, assuming that one
of the operands needs to be fetched from the memory indicated.  Fetching one operand essentially assumes that the operations perfectly cascade, so the output of the operation is stored into the register file and then read out as the other operand for the next operation.  For simple 16-bit operations, accesses to even a small memory are very costly
(10x GOPS/W when going from {\it Mult} to {\it 4K~SRAM} in the table),
while for more expensive 64-bit operations, first level cache accesses only triple the energy cost
(from about 45 GOPS/W {\it Mult} to 15 GOPS/W {\it 4K~SRAM}).  
For 64-bit FP, it is the last level cache and DRAM accesses that have a dramatic effect.
It is important to remember that this limitation is independent of the degree of parallelism of the
application or the hardware.  For memory, parallelism does not change
the energy/access, and thus does not change the peak bandwidth in a
power limited system. 

\subsection{Emerging Memory Technologies}
Recently there has been an increasing interest in new memory technologies
fueled by the possibility that more radical developments in memory or
interconnect technology 
will emerge.
%
%
Examples of these technological changes include increasing on-die memory using existing or emerging technologies such as 
eDRAM~\cite{eDRAM}, 
STT-MRAM~\cite{STT-MRAM1}, 
RRAM~\cite{RRAM}, 
PCM~\cite{PCM} or 
3D Xpoint~\cite{merritt2016xpoint}, to using RRAM, PCM or Xpoint to replace DRAM or adding an additional level after DRAM in the memory hierarchy. 
Most of these technologies are non-volatile so have a low leakage state, and can be stacked to yield very high densities. 
These new technologies 
are proposed for creating large memories, 
and these large memories will need long wires 
to distribute the address and data.  Thus, while the length of these wires might be shorter than in DRAM, they will still be long enough to require significant energy compared to computation, and 
must be used
infrequently.
Hence, the need for dark memory is an inherent issue in the design of the system for any reasonable memory solution in the foreseeable future.

Given the criticality of keeping the memory 
hierarchy---especially the DRAM---dark, 
the first part of accelerator
design is not about the hardware: it is to find 
a way to execute the application using an algorithm that minimizes DRAM accesses and has high chip-level locality, especially when parallelized, as described in the next section.

\section{Algorithmic Optimization}
\label{sec:codesign}

Given the high cost of memory accesses, algorithm optimization primarily focuses on minimizing DRAM and low-level cache accesses, and secondarily creating parallelism that can be exploited on chip.  The simplest optimizations involve blocking, which splits and reorders loops to increase locality.  In this context, it is possible to unroll a loop in hardware, creating parallelism for the hardware to exploit.  Often these methods are not enough, however, and a new lower-communication approach to the problem is needed.  
That approach
can have a higher computation cost, but if the energy is communication dominated, 
it is
still more energy efficient.

\subsection{Exploiting Locality and Blocking}

\begin{figure}[tb] 
   \centering
   \includegraphics
     [width=\columnwidth]{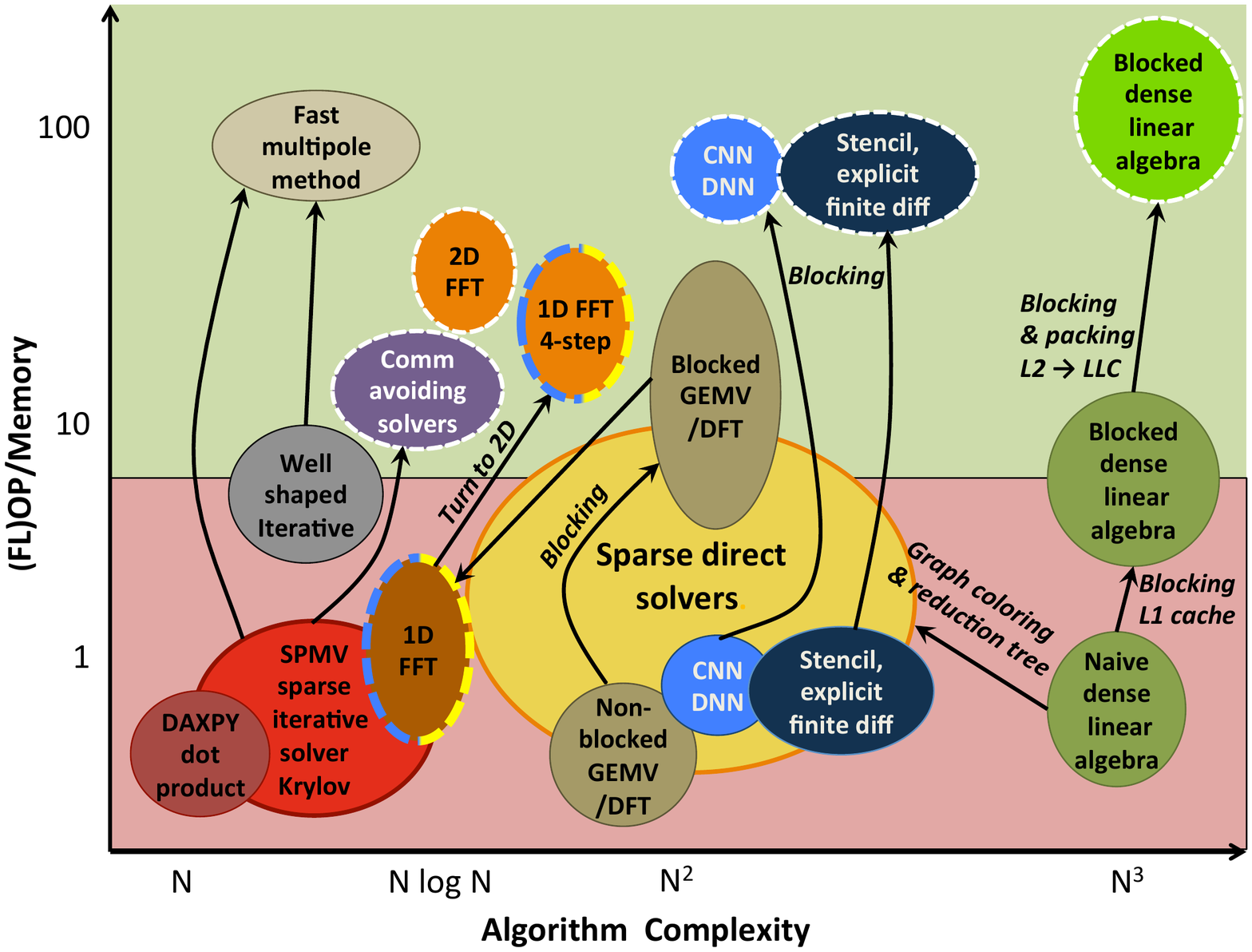} 
   \vspace{-\baselineskip}
      \vspace{-10pt}
   \caption{Complexity vs. computation/memory-access ratio for several algorithms. Dashed algorithms increase algorithm complexity for efficient implementation.}
   \vspace{-10pt}
   \label{fig:Apps}
\end{figure}

We will use GEneral Matrix Multiplication (GEMM) $A \times B = C$ as an example to see how blocking can reduce DRAM accesses and consequently save energy.\footnote{As part of the BLAS scientific computing library, GEMM is essential to innumerable applications, 
including data parallel applications.} At first GEMM looks like it should be computation dominated, since for $n$ by $n$ matrices it accesses $3n^2$ memory locations (read two and write one matrix) and performs $2n^3$ operations. The problem arises with the required working set of a naive implementation, since to create one row of the output requires reading the entire $B$ matrix, which can be very large.  As a result this matrix must be reread $n$ times, leading to $n^3$ memory operations and
low FLOPS per DRAM access as depicted in Figure~\ref{fig:Apps} 
(as ``naive dense linear algebra").

However by reordering the computation, we can greatly increase the locality. If we view each matrix as composed of a number of smaller $ b \times b$ matrices, each entire sub-matrix can be stored in a $b\times b$ block of memory on-chip. Now if we iterate over these sub-matrices we need to refetch the $B$ matrix only $n/b$ times, reducing the DRAM accesses down to $2n^3/b + 2n^2$ accesses~\cite{lam1991cache}.  This technique can be applied recursively, blocking each sub-matrix into a higher level of the memory hierarchy, with the highest level blocked into the register file. Adding this on-chip memory increases the area and power dissipated by the chip, but causes the system power to greatly decrease by keeping the DRAM dark. 
As Figure~\ref{fig:Apps} demonstrates, 
blocking can improve many computations, including 
algorithms for dense linear 
algebra~\cite{lam1991cache,renganarayana2004tiling,navarro1994blocking}, convolutional neural networks~\cite{chen2014dadiannao},
the four-step FFT~\cite{vanloan1992fft,bailey1989ffts,takahashi2000fft} and many others.

\subsection{Sequential to Parallel}
Locality is also 
critical
when mapping an application to parallel hardware, since it is best if the parallel executions use mostly local data.  Both data and task parallelism can be exploited in hardware design, which often requires small algorithmic changes to remove minor dependencies in the sequential code. Data parallelism is often exploited by taking one of the blocked loops and unrolling it so each loop iteration is done by a different piece of hardware, while task parallelism is exploited by building a hardware block for each task, and using wires to handle the producer/consumer communication. 

Parallel execution generally decreases the energy required for memory that is strictly local to the unit, since in this case the original memory is partitioned into many smaller memories with one memory embedded into each parallel unit. The energy required for memory storing shared data generally goes up, since now this data must be communicated to all the cores, which are large in size due to their private memory.  We will again use GEMM to demonstrate this issue. To create a parallel GEMM execution, we distribute the rows of A to different cores and broadcast the columns of B to all the cores so each core produces unique rows of C. Since the A and C matrix are partitioned among the cores, the working set in each core is smaller, since it only needs to hold a fraction of the total matrix. The memory required for the B matrix remains the same size, but now its output needs to be broadcast to all the cores~\cite{pedram2012codesign}. The energy required to distribute this information is proportional to the square root of the area that all the cores occupy, which is related to the total memory used in all the cores (plus the overhead of the hardware), and is often larger than the energy needed to fetch B from its memory.  This overhead makes it critical for parallel algorithms to limit the total communication between parallel units, or restrict them to physically adjacent units.

\subsection{Changing the nature of the algorithm}
While it may be possible to get the required locality and parallelism through blocking, sometimes a very different approach is needed to reach the desired performance.  Here the application developer needs to take broader look at the problem, to see if there are problem symmetries or simplifications that can be exploited, different approaches to try, or constraints that can be relaxed. 
For example, in linear algebra,
different variants of algorithms 
show different behaviors in various levels of the memory hierarchy
so
the specific choice of
variant affects locality and 
performance~\cite{bientinesi2005science,anderson1990evaluating}.
Another example is 
the FFT, which exploits symmetries in the DFT to dramatically reduce the complexity of computing a Fourier transform~\cite{cooley1965algorithm}.  

A classical example depicted in Figure~\ref{fig:Apps} is the solution of sparse systems. The most straightforward method is to use expensive O($N^3$)
dense direct methods that do not take advantage of sparsity in the data structure. 
Sparse direct solvers use techniques like reordering the data, graph coloring~\cite{naumovparallel}, and constructing dependence trees to preserve non-zero patterns in the matrix and so avoid performing computations with zeros, all while improving parallelism~\cite{heath1991parallel}.
This drops the computations\footnote{For matrices whose graphs can be embedded in at most three dimensions.} down to at most O($N^2$) in spite of various overheads for extra complexity. 
In contrast, iterative solvers reduce computations by performing a sequence of improving approximate solutions that are much cheaper in complexity~(e.g.~O($N^2$)) 
and~(for well-conditioned matrices)~converge after a few iterations~\cite{barrett1994templates}. However, each iteration consists of low-performance memory-bound kernels like (sparse)~matrix-vector multiplication. Communication-avoiding algorithms can replace these memory-bound kernels with GEMM like kernels to improve the locality and performance at the cost of slightly slower convergence rate and more computations~\cite{hoemmen2010communication,demmel2012communication}.


Other approaches relax some constraints in the original problem.  For example, 
iterative refinement techniques use high precision arithmetic for lower order residual computation
and then use lower precision arithmetic for 
high order less-sensitive linear solve kernels~\cite{wilkinson1994rounding}.
This method 
can speed the computation by up to two orders of magnitude
and can be generalized for solving linear least square problems, eigenvalue/singular value computations, and sparse solutions like conjugate gradient~\cite{langou2006exploiting}.
Or parallel applications can allow cores that update shared state to be stochastic with respect to other processors.  Both of these methods 
sacrifice convergence rate to decrease communication for each computation round.

This reduction of constraints is widely used in applications that use
randomized algorithms, which are becoming 
popular especially in domains like machine learning and 
Principal Components Analysis (PCA)
where approximate but fast results are desired. Such methods select a random subset of the initial input data and reduce substantial parts of the computation while still managing to converge on a desired result~\cite{scholkopf1999pca,moore1981pca,FLAWN78}.

 \section{Metrics for energy constrained computing}
\label{sec:metrics}

\begin{figure}
    \centering
    \includegraphics[
      width=1\columnwidth
    ]{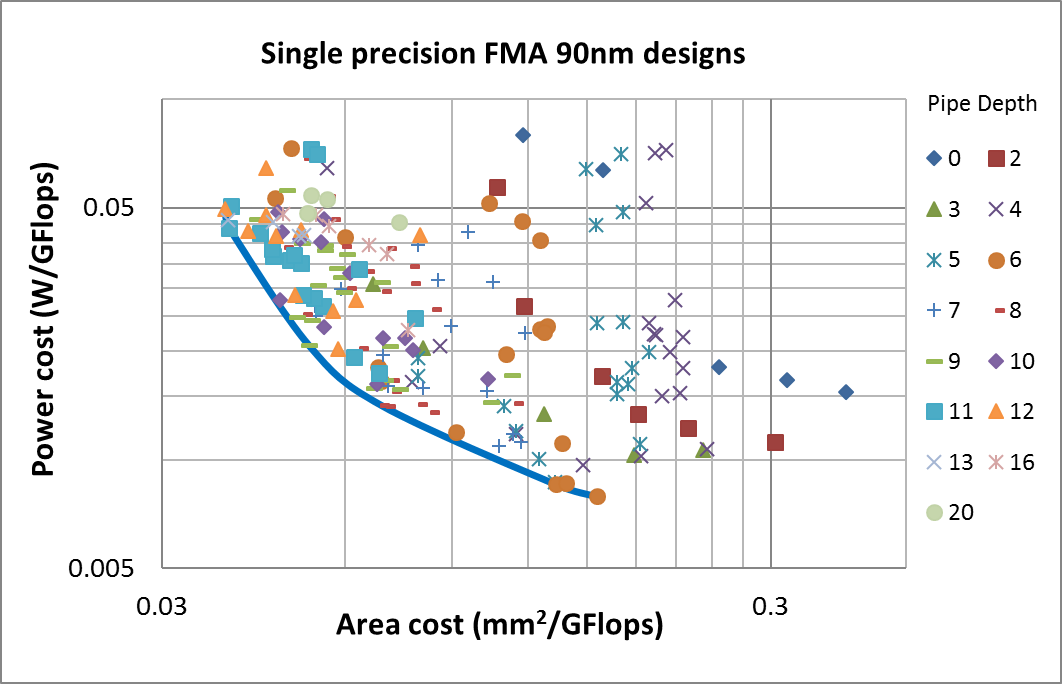}
       \vspace{-20pt}
    \caption{Mapping of a large design space of fused FMADD designs.
    Each dot represents a different variation on the base design; for example, all the diamond shapes represent various unpipelined versions, squares have a pipe depth of 2 and so on up to a 20-deep pipeline.
    Most of the designs are strictly worse 
    in the sense that they either take more area or more energy than one of the other designs. The left-hand edge is the edge of the feasible space, and these designs are optimal for some design constraints~\cite{galal2011energy}.}
    \label{fig:power_area_FMA_design_Space}
\end{figure}

To formalize the trade-offs discussed in the previous sections we will assume that we are building 
a system on chip (SoC) 
with specialized hardware designed to solve a data parallel problem,
and that we have constraints on, or want to optimize combinations of performance, power and chip area.\footnote{Talk of free transistors aside, die area is still 
important to consider. It strongly affects cost when you sell parts in large ($10^6$) volumes, and low volume parts still have area constraints they can't exceed.} To solve this optimization problem we can place every possible design combination in a 3-D space, where the x-axis is chip area, the y-axis is power, and the z-axis is performance.  In this space it is easy to remove designs that can never be optimal: designs with the same area and power as another design but lower performance,  designs with the same performance and area but higher energy, or designs with the same performance and energy by larger area.  Removing these suboptimal designs will leave a 2-D surface of designs that might be optimal.

Fortunately we can simplify this space further by recognizing that we are solving a data parallel problem.
In this type of problem we assume 
you can double the throughput (the performance) by doubling the hardware (the power and area).  What this means is that each design is not a {\it point} in the 3-D performance space, but a line. To convert a design back to a point, we divide the area and power axes by the performance of the design (since both of these parameters are proportional to performance) and end up with a 2-D metric space: power/performance, or energy/op; versus area/performance, or mm$^2$/op/s.

\subsection{Joules/op and mm$^2$/(ops/s) Metrics}

\begin{figure}[t]
	\centering
	\subfigure[]{\includegraphics[width=0.95\columnwidth]
	{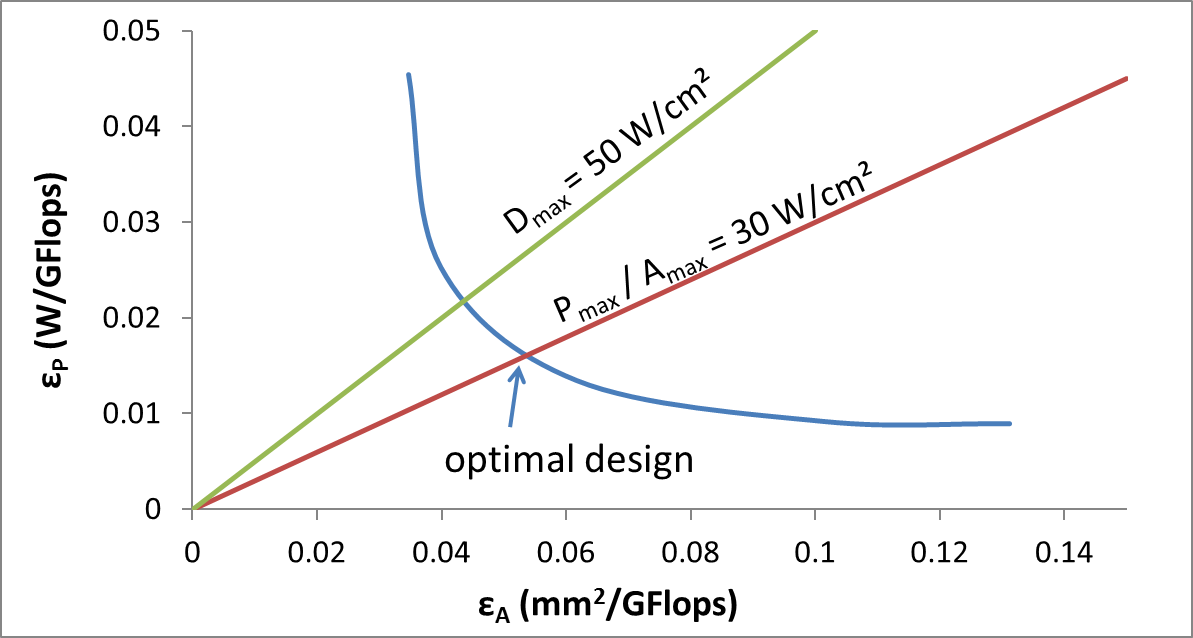}}
	\subfigure[]{\includegraphics[width=0.96\columnwidth]
	{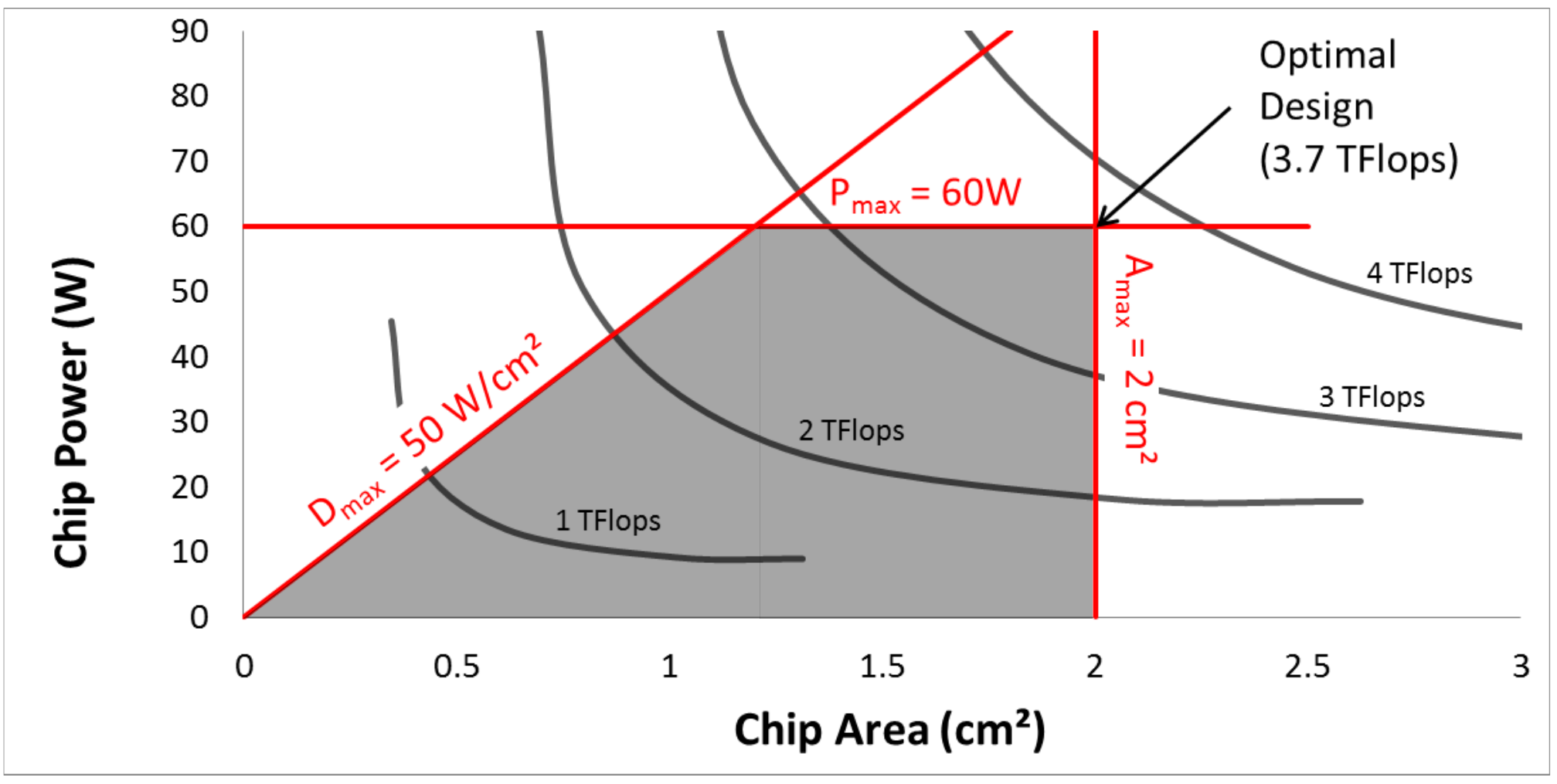}}
	\vspace{-5pt}
	\caption[Determining Optimal Resource Constrained Design]{(a) Determining the optimal design point from a throughput-energy trade-off curve and constraints. (b) Contour map of achievable throughputs versus area and power. Constraints of A\textsubscript{max} = 2 cm\textsuperscript{2}, P\textsubscript{max} = 60 W, and D\textsubscript{max} = 50 W/cm\textsuperscript{2} are indicated~\cite{galal2012thesis}.}
	\vspace{-15pt}
	\label{fig:ResourceConstrainedThroughput}
\end{figure}

Like in the 3-D case, it is easy to find non-optimal designs. Any design that has a higher energy/op with the same compute density as another design can never be the best design.  
Similarly if two designs have the same energy/op, the one with a higher mm$^2$/(ops/s) cost cannot be optimal.  
Figure~\ref{fig:power_area_FMA_design_Space}
%
shows the result of 
evaluating the design space for an FP fused mult-add unit, and exploring different microarchitecture, pipeline depth, gate sizing, cell libraries, and Vdd 
settings.
From an energy efficient design perspective, we can completely
characterize this design space, which includes the effect of dimming, by the shape of its Pareto curve 
(the left hand edge of the feasible space), 
which is shown in Figure~\ref{fig:ResourceConstrainedThroughput}a.

These two metrics nicely capture many of the trade-offs we have discussed previously.  As we \textit{dim} the silicon, we create designs with lower energy/op, but they will also operate slower, which moves along the Pareto curve. Similarly adding a level in the memory hierarchy may decrease the energy of an access, but will also increase the area required, 
contributing another design point to the Pareto curve.

To show why Figure~\ref{fig:ResourceConstrainedThroughput}a is so powerful,  Figure~\ref{fig:ResourceConstrainedThroughput}b plots the power and area of an accelerator,
and shows some possible design constraints.  Notice that the lines of constant performance
shown in this plot are simply 
the Figure~\ref{fig:ResourceConstrainedThroughput}a curve
scaled by different throughput numbers.  So finding the maximum performance point for $P$\,$<$\,$P_{max}$ and 
$A$\,$<$\,$A_{max}$
is the same as finding
%
%
the point 
$\big(${\large $\varepsilon$}$_A$,\,{\large $\varepsilon$}$_P\big)$ 
in Figure~\ref{fig:ResourceConstrainedThroughput}a where
{\small $($energy/op$)$}\,/\,{\small $($mm$^2$/(ops/s)$)$} = $P_{max}/A_{max}$,
and the resulting performance is 
$A_{max}$\,/\,{\large $\varepsilon$}$_A$. 
Other optimization objectives can be mapped to a curve in this space, allowing them to be solved
as well, including optimizing for total cost of ownership.  For more details 
see Galal's work on energy-efficient FPU design~\cite{galal2011energy}.

If the algorithm is fixed, one can use any definition of an op in these metrics, 
since this optimization does not change the number of ops.
However, if we need to compare designs across different algorithmic approaches, it is essential to define {\it op} to be something that is invariant across the different implementations.  For example 
using FLOPs
to compare sparse and dense algorithms would be a bad idea, since a dense implementation would have much lower energy/FLOP and area/FLOP/s, but would require many more FLOPs than a sparse solver, and would 
look worse
on the curve.  Similarly, when trading-off among different possible implementations, it is important that they all use the same op definition.

\subsection{Accelerator Optimization}
\label{sec:how_to_use_pareto_curves}


Another advantage of using Pareto curves rather than a specific design point is that the curve provides you information about marginal cost in area or energy if you need to change the design.  While these marginal costs assume you can add fractional compute units to get fractional performance, which is clearly wrong, they do provide the insight needed to create efficient solutions. To demonstrate how they can be used for accelerator evaluation, assume  
our application is
running on a scalable machine and we
want to minimize
this machine's power by adding some specialized 
accelerators while staying within the
chip's current area and performance budget. Since we are assuming the base machine and accelerator area scale with performance, moving computation from the base machine to the accelerator will provide area that the accelerator can use.  The accelerator will improve the energy of the machine if it has a lower energy/op when operating at the same mm$^2$/(op/s) as 
the base machine.
Since the compute density is the same, this new solution should require the same area as before.

The previous step verified that the accelerator can reduce energy/op versus the original system, but the resulting design is not necessarily optimal: to ease the comparison we chose points that had the same compute density, and left the base design alone.  We need to change both to get the optimal power.  Fortunately, like most constrained optimization problems, the optimal area allocation can be found by balancing marginal costs: at the optimal point, the change in energy/op per change in mm$^2$/(op/s) in the two compute units must be the same. Moving an increment of work lowers the energy of the unit losing the work by its marginal cost, while the unit gaining the work increases its energy by its marginal cost. If these are not the same, moving work from the unit with higher marginal cost to the one with lower marginal cost will save energy (or if the work cannot move because the accelerator is specialized, simply move silicon area in the other direction).

\subsection{Non-Scalable Objects}
\label{sec:non_scalable}

While this method clearly shows how Pareto information 
lets us
optimally allocate area between two compute engines,
its assumption of finely partitioned engines is rarely the case.
In most designs, the area of a block cannot be smoothly changed.  Processors/accelerators can be scaled
by duplication, but since each unit contains 
compute/control/memory they are generally of significant
size. The result is one cannot really incrementally move 
area from one unit to the other.  Instead you can only make
much coarser-grain moves.  This quantization makes finding 
the exact answer harder, since now we need to solve a mixed 
integer program; but the basic intuition remains the same: 
%
%
If the marginal cost of an accelerator $A_1$ is lower than a second unit $A_2$,
test to see if you can reduce the size of $A_1$
enough to give $A_2$ enough area so it can move to a more energy efficient design. This 
might involve
lowering the performance of 
each existing $A_2$ compute unit, and then adding a new one
to maintain aggregate throughput. If enough area cannot be created, the best alternative is to try to use the area in $A_1$ to reduce its energy cost.

Dealing with the memory system 
adds a new
challenge.  While the register files and first level caches are duplicated with the compute units, the levels in the memory hierarchy closer to DRAM (last level cache, and sometimes even the L2) are shared and so their area is not proportional to the computing throughput. Fortunately, like a compute unit, one can create a Pareto curve for a memory system.  The y-axis remains energy/op, but now it represents the
average memory energy used for each processor op.  Since area does not scale with performance the
x-axis is just area.
Like compute units, the different memory configurations will collectively generate a single Pareto curve, where larger area
reduces the average memory cost, by filtering out more of the DRAM accesses. 

This memory Pareto curve has exactly the form we need to find the optimal allocation between memory and computation. 
We just
scale the  compute curve by the desired aggregate performance so its Pareto
curves also indicate the trade-off between area and energy/op, and the energy optimal design
will balance the marginal cost between the two units.

\begin{figure}
	\centering
	\vspace{-5pt}
	\subfigure[]{\includegraphics[width=0.96\columnwidth]
	{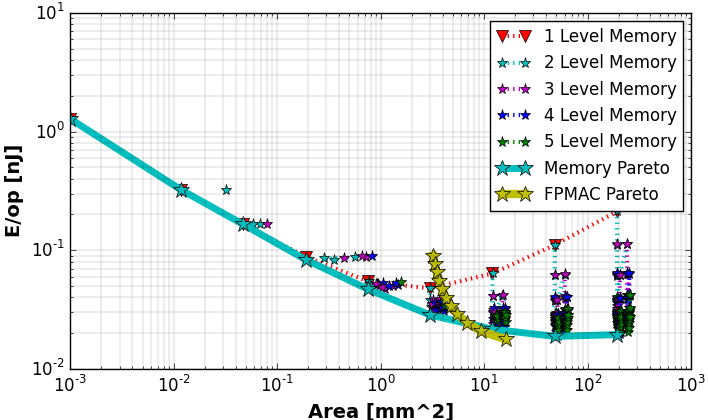}}
	\subfigure[]{\includegraphics
	[width=.95\columnwidth,clip,trim=8pt 8pt 16pt 8pt] 
	{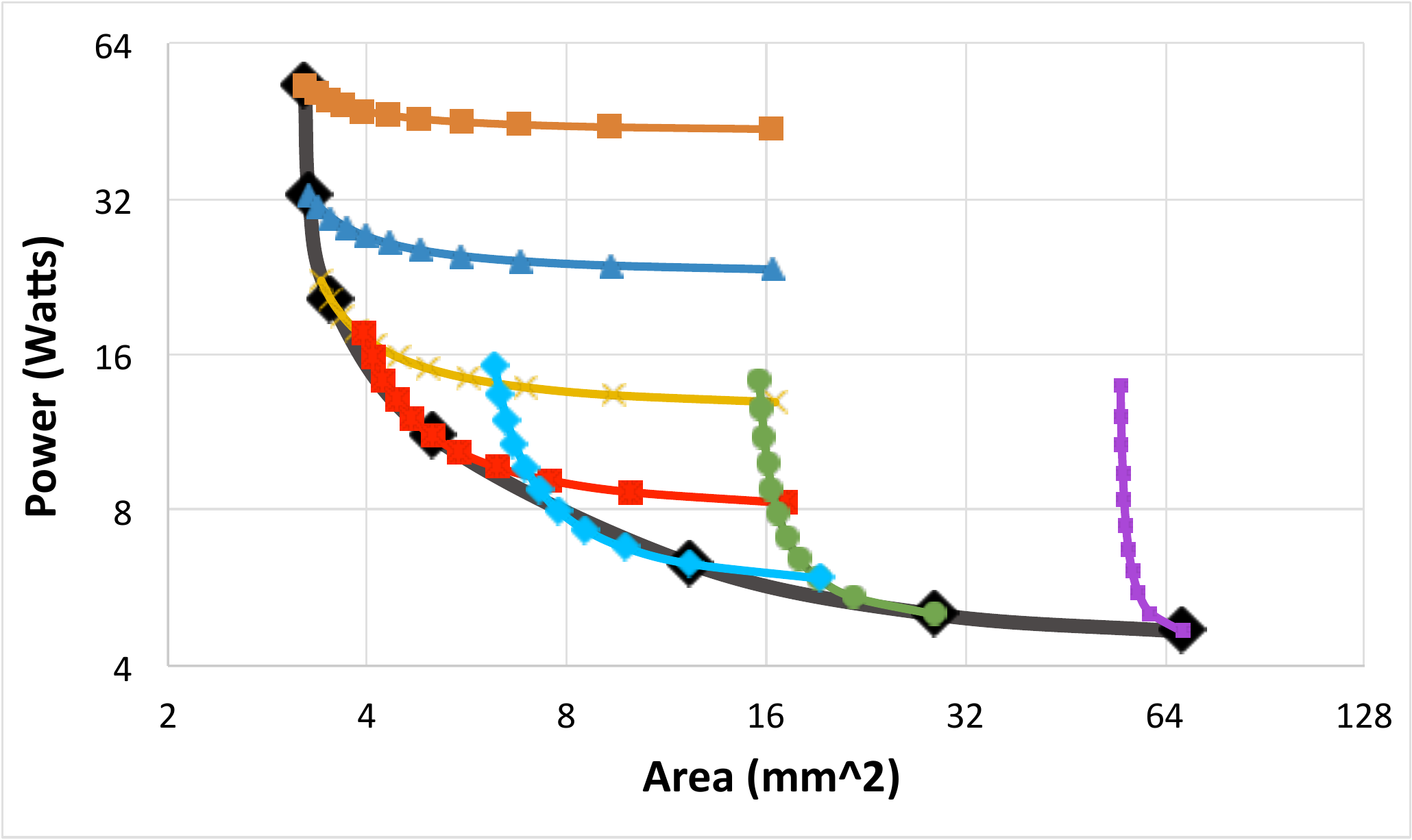}}	
	\caption[Pareto Representation of Memory, FP-MAC, and Architecture design space for GEMM]{(a) Effect of multiple layers of on-chip memory on the energy and area tradeoffs for GEMM. As the area grows, more memory levels are needed in the optimal design.  One level memory is registers and DRAM, two levels has registers, local memory 
	and DRAM, etc. (b) Pareto curve of 256-GFLOP GEMM accelerator, shown in black.  This was generated by finding the FMADD design that matched the margin cost of the memory system.  Also drawn are the systems that would result by pairing different FMADD trade-off choices to the optimal memory design points showing other potential designs, most of which are highly sub-optimal. 
	}
	\label{fig:GEMM_Example}
\end{figure}

Figure~\ref{fig:GEMM_Example} shows how this is done for a GEMM accelerator.  Using the known access pattern of the algorithm, the required memory energy per fused multiply/add is found for all possible memory configurations.  We explore 1-5 levels of on-chip memory hierarchy in addition to the DRAM, and try many different potential memory sizes for each level. Most of these configurations are not optimal, but a few form the Pareto curve (in turquoise). This curve shows how the memory energy changes from 1\,nJ/FMADD to around 20\,pJ/FMADD as the area changes from 0 to 100mm$^2$.  Also shown in Figure~\ref{fig:GEMM_Example}(a) is the Pareto curve of an FMADD running at 256~GFLOPS.  
To generate the power and area curve for the entire system, we add the energy and area cost of the FMADD design at each point in the memory Pareto curve.  This results in the many curves shown in Figure~\ref{fig:GEMM_Example}(b). Overlayed on these curves is the overall Pareto curve, which is shown in black which uses the FMADD design which matches the marginal cost of the memory system. Not surprisingly, the small area solutions chose high compute density FMADD solutions, since the memory system dominates the energy, while large memory area solutions use low energy, and area inefficient FMADD. The result is that even though the total power ranges by nearly 10x, in most of these designs the compute energy and memory energy are roughly 50/50. 


\section{Conclusion}
The large energy cost of memory fetches limits the overall efficiency of applications no matter how efficient the accelerators are on the chip. As a result the most important optimization must be done at the algorithm level, to reduce off-chip memory accesses, to create {\it Dark Memory}.
The algorithms must first be (re)written for both locality and parallelism before you tailor the hardware to accelerate them.

Using Pareto curves in the {\it energy/op} and {\it mm$^2$/(op/s)} space allows one to quickly evaluate different accelerators, memory systems, and even algorithms to understand the trade-offs between performance, power and die area. This analysis is a powerful way to optimize chips in the Dark Silicon era.

\bibliographystyle{IEEEtran/IEEEtran}
\bibliography{bib/ref.bib}

\begin{thebibliography}{10}
\providecommand{\url}[1]{#1}
\csname url@samestyle\endcsname
\providecommand{\newblock}{\relax}
\providecommand{\bibinfo}[2]{#2}
\providecommand{\BIBentrySTDinterwordspacing}{\spaceskip=0pt\relax}
\providecommand{\BIBentryALTinterwordstretchfactor}{4}
\providecommand{\BIBentryALTinterwordspacing}{\spaceskip=\fontdimen2\font plus
\BIBentryALTinterwordstretchfactor\fontdimen3\font minus
  \fontdimen4\font\relax}
\providecommand{\BIBforeignlanguage}[2]{{%
\expandafter\ifx\csname l@#1\endcsname\relax
\typeout{** WARNING: IEEEtran.bst: No hyphenation pattern has been}%
\typeout{** loaded for the language `#1'. Using the pattern for}%
\typeout{** the default language instead.}%
\else
\language=\csname l@#1\endcsname
\fi
#2}}
\providecommand{\BIBdecl}{\relax}
\BIBdecl

\bibitem{Dennard:jssc74}
R.~H. Dennard, F.~H. Gaensslen, H.~N. Yu, V.~L. Rideout, E.~Bassous, and A.~R.
  LeBlanc, ``Design of ion-implanted {MOSFET}s with very small physical
  dimensions,'' \emph{IEEE Journal of Solid-State Circuits}, vol.~9, no.~5, pp.
  256 -- 268, 1974.

\bibitem{danowitz2012cpu}
A.~Danowitz, K.~Kelley, J.~Mao, J.~P. Stevenson, and M.~Horowitz, ``{CPU DB}:
  recording microprocessor history,'' \emph{Communications of the ACM},
  vol.~55, no.~4, pp. 55--63, 2012.

\bibitem{esmaeilzadeh2011dark}
H.~Esmaeilzadeh, E.~Blem, R.~S. Amant, K.~Sankaralingam, and D.~Burger, ``Dark
  silicon and the end of multicore scaling,'' in \emph{Computer Architecture
  (ISCA), 2011 38th Annual International Symposium on}.\hskip 1em plus 0.5em
  minus 0.4em\relax IEEE, 2011, pp. 365--376.

\bibitem{taylor2012dark}
M.~B. Taylor, ``Is dark silicon useful?: harnessing the four horsemen of the
  coming dark silicon apocalypse,'' in \emph{Proceedings of the 49th Annual
  Design Automation Conference}.\hskip 1em plus 0.5em minus 0.4em\relax ACM,
  2012, pp. 1131--1136.

\bibitem{grenat20145}
A.~Grenat, S.~Pant, R.~Rachala, and S.~Naffziger, ``Adaptive clocking system
  for improved power efficiency in a 28nm x86-64 microprocessor,'' in
  \emph{Solid-State Circuits Conference Digest of Technical Papers (ISSCC),
  2014 IEEE International}.\hskip 1em plus 0.5em minus 0.4em\relax IEEE, 2014,
  pp. 106--107.

\bibitem{horowitz2014keynote}
M.~Horowitz, ``Computing's energy problem (and what we can do about it),'' in
  \emph{Solid-State Circuits Conference Digest of Technical Papers (ISSCC),
  2014 IEEE International}, Feb 2014, pp. 10--14.

\bibitem{galal2012thesis}
S.~Galal, ``Energy efficient floating-point unit design,'' Ph.D. dissertation,
  Stanford University, 2012.

\bibitem{amrutur2000speed}
B.~S. Amrutur and M.~A. Horowitz, ``Speed and power scaling of {SRAM}'s,''
  \emph{Solid-State Circuits, IEEE Journal of}, vol.~35, no.~2, pp. 175--185,
  2000.

\bibitem{evans1995energy}
R.~J. Evans and P.~D. Franzon, ``Energy consumption modeling and optimization
  for {SRAM}'s,'' \emph{Solid-State Circuits, IEEE Journal of}, vol.~30, no.~5,
  pp. 571--579, 1995.

\bibitem{eDRAM}
S.~Narasimha, P.~Chang, C.~Ortolland, D.~Fried, E.~Engbrecht, K.~Nummy,
  P.~Parries, T.~Ando, M.~Aquilino, N.~Arnold, R.~Bolam, J.~Cai, M.~Chudzik,
  B.~Cipriany, G.~Costrini, M.~Dai, J.~Dechene, C.~DeWan, B.~Engel,
  M.~Gribelyuk, D.~Guo, G.~Han, N.~Habib, J.~Holt, D.~Ioannou, B.~Jagannathan,
  D.~Jaeger, J.~Johnson, W.~Kong, J.~Koshy, R.~Krishnan, A.~Kumar, M.~Kumar,
  J.~Lee, X.~Li, C.~Lin, B.~Linder, S.~Lucarini, N.~Lustig, P.~McLaughlin,
  K.~Onishi, V.~Ontalus, R.~Robison, C.~Sheraw, M.~Stoker, A.~Thomas, G.~Wang,
  R.~Wise, L.~Zhuang, G.~Freeman, J.~Gill, E.~Maciejewski, R.~Malik, J.~Norum,
  and P.~Agnello, ``22nm high-performance {SOI} technology featuring
  dual-embedded stressors, epi-plate high-{K} deep-trench embedded {DRAM} and
  self-aligned via {15LM BEOL},'' in \emph{Electron Devices Meeting (IEDM),
  2012 IEEE International}, Dec 2012, pp. 3.3.1--3.3.4.

\bibitem{STT-MRAM1}
H.~Yoda, S.~Fujita, N.~Shimomura, E.~Kitagawa, K.~Abe, K.~Nomura, H.~Noguchi,
  and J.~Ito, ``Progress of {STT-MRAM} technology and the effect on
  normally-off computing systems,'' in \emph{Electron Devices Meeting (IEDM),
  2012 IEEE International}, Dec 2012, pp. 11.3.1--11.3.4.

\bibitem{RRAM}
H.-S. Wong, H.-Y. Lee, S.~Yu, Y.-S. Chen, Y.~Wu, P.-S. Chen, B.~Lee, F.~Chen,
  and M.-J. Tsai, ``Metal oxide {RRAM},'' \emph{Proceedings of the IEEE}, vol.
  100, no.~6, pp. 1951--1970, June 2012.

\bibitem{PCM}
H.-S. Wong, S.~Raoux, S.~Kim, J.~Liang, J.~P. Reifenberg, B.~Rajendran,
  M.~Asheghi, and K.~E. Goodson, ``Phase change memory,'' \emph{Proceedings of
  the IEEE}, vol.~98, no.~12, pp. 2201--2227, Dec 2010.

\bibitem{merritt2016xpoint}
R.~Merritt, ``{3D} {XP}oint steps into the light,'' \emph{EE Times}, 14 Jan
  2016.

\bibitem{lam1991cache}
M.~D. Lam, E.~E. Rothberg, and M.~E. Wolf, ``The cache performance and
  optimization of blocked algorithms,'' \emph{ACM SIGOPS Operating Systems
  Review}, vol.~25, no. Special Issue, pp. 63--74, 1991.

\bibitem{renganarayana2004tiling}
L.~Renganarayana and S.~Rajopadhye, ``A geometric programming framework for
  optimal multi-level tiling,'' in \emph{Proceedings of the 2004 ACM/IEEE
  Conference on Supercomputing}, ser. SC '04.\hskip 1em plus 0.5em minus
  0.4em\relax Washington, DC, USA: IEEE Computer Society, 2004, p.~18.

\bibitem{navarro1994blocking}
J.~J. Navarro, T.~Juan, and T.~Lang, ``{MOB} forms: A class of multilevel block
  algorithms for dense linear algebra operations,'' in \emph{Proceedings of the
  8th International Conference on Supercomputing}, ser. ICS '94.\hskip 1em plus
  0.5em minus 0.4em\relax New York, NY, USA: ACM, 1994, pp. 354--363.

\bibitem{chen2014dadiannao}
Y.~Chen, T.~Luo, S.~Liu, S.~Zhang, L.~He, J.~Wang, L.~Li, T.~Chen, Z.~Xu,
  N.~Sun, and O.~Temam, ``{DaDianNao:} a machine-learning supercomputer,'' in
  \emph{47th Annual IEEE/ACM Int'l Symp. on Microarchitecture (MICRO)}.\hskip
  1em plus 0.5em minus 0.4em\relax IEEE, 2014, pp. 609--622.

\bibitem{vanloan1992fft}
C.~Van~Loan, \emph{Computational Frameworks for the Fast Fourier
  Transform}.\hskip 1em plus 0.5em minus 0.4em\relax Society for Industrial and
  Applied Mathematics, 1992.

\bibitem{bailey1989ffts}
D.~H. Bailey, ``{FFT}s in external or hierarchical memory,'' in \emph{Proc.
  1989 ACM/IEEE conference on Supercomputing}.\hskip 1em plus 0.5em minus
  0.4em\relax ACM, 1989, pp. 234--242.

\bibitem{takahashi2000fft}
D.~Takahashi, ``High-performance parallel {FFT} algorithms for the {H}itachi
  {SR}8000,'' in \emph{High Performance Computing in the Asia-Pacific Region,
  2000. Proceedings. The Fourth International Conference/Exhibition on},
  vol.~1, May 2000, pp. 192--199 vol.1.

\bibitem{pedram2012codesign}
A.~Pedram, R.~Van~de Geijn, and A.~Gerstlauer, ``Codesign tradeoffs for
  high-performance, low-power linear algebra architectures,'' \emph{Computers,
  IEEE Transactions on}, vol.~61, no.~12, pp. 1724--1736, 2012.

\bibitem{bientinesi2005science}
P.~Bientinesi, J.~A. Gunnels, M.~E. Myers, E.~S. Quintana-Ort{\'\i}, and R.~A.
  van~de Geijn, ``The science of deriving dense linear algebra algorithms,''
  \emph{ACM Trans. Mathematical Software (TOMS)}, vol.~31, no.~1, pp. 1--26,
  2005.

\bibitem{anderson1990evaluating}
E.~Anderson and J.~Dongarra, \emph{Evaluating Block Algorithm Variants in
  LAPACK.}\hskip 1em plus 0.5em minus 0.4em\relax University of Tennessee Dept.
  of Computer Science, 1990.

\bibitem{cooley1965algorithm}
J.~W. Cooley and J.~W. Tukey, ``An algorithm for the machine calculation of
  complex {F}ourier series,'' \emph{Math. comput}, vol.~19, no.~90, p. 297,
  1965.

\bibitem{naumovparallel}
M.~Naumov, P.~Castonguay, and J.~Cohen, ``Parallel graph coloring with
  applications to the incomplete-{LU} factorization on the {GPU},'' Nvidia
  Corp., Tech. Rep. NVR-2015-001, 2015.

\bibitem{heath1991parallel}
M.~T. Heath, E.~Ng, and B.~W. Peyton, ``Parallel algorithms for sparse linear
  systems,'' \emph{SIAM review}, vol.~33, no.~3, pp. 420--460, 1991.

\bibitem{barrett1994templates}
R.~Barrett, M.~W. Berry, T.~F. Chan, J.~Demmel, J.~Donato, J.~Dongarra,
  V.~Eijkhout, R.~Pozo, C.~Romine, and H.~Van~der Vorst, \emph{Templates for
  the solution of linear systems: building blocks for iterative methods}.\hskip
  1em plus 0.5em minus 0.4em\relax Siam, 1994, vol.~43.

\bibitem{hoemmen2010communication}
M.~Hoemmen, ``Communication-avoiding {K}rylov subspace methods,'' Ph.D.
  dissertation, University of California, Berkeley, 2010.

\bibitem{demmel2012communication}
J.~Demmel, L.~Grigori, M.~Hoemmen, and J.~Langou, ``Communication-optimal
  parallel and sequential {QR} and {LU} factorizations,'' \emph{SIAM Journal on
  Scientific Computing}, vol.~34, no.~1, pp. A206--A239, 2012.

\bibitem{wilkinson1994rounding}
J.~H. Wilkinson, \emph{Rounding errors in algebraic processes}.\hskip 1em plus
  0.5em minus 0.4em\relax Courier Corporation, 1994.

\bibitem{langou2006exploiting}
J.~Langou, P.~Luszczek, J.~Kurzak, A.~Buttari, and J.~Dongarra, ``Exploiting
  the performance of 32 bit floating point arithmetic in obtaining 64 bit
  accuracy (revisiting iterative refinement for linear systems),'' in
  \emph{Proc. ACM/IEEE SC 2006 Conf.}\hskip 1em plus 0.5em minus 0.4em\relax
  IEEE, 2006, pp. 50--50.

\bibitem{scholkopf1999pca}
B.~Scholkopf, S.~Mika, C.~Burges, P.~Knirsch, K.~Muller, G.~Ratsch, and
  A.~Smola, ``Input space versus feature space in kernel-based methods,''
  \emph{IEEE Trans. Neural Networks}, vol.~10, no.~5, pp. 1000--1017, Sep 1999.

\bibitem{moore1981pca}
B.~Moore, ``Principal component analysis in linear systems: Controllability,
  observability, and model reduction,'' \emph{Automatic Control, IEEE
  Transactions on}, vol.~26, no.~1, pp. 17--32, Feb 1981.

\bibitem{FLAWN78}
P.-G. Martinsson, G.~Quintana-Orti, N.~Heavner, and R.~van~de Geijn,
  ``Householder {QR} factorization: Adding randomization for column pivoting,''
  The University of Texas at Austin, Department of Computer Sciences, Technical
  Report {FLAME} {W}orking {N}ote \#78, December 2015.

\bibitem{galal2011energy}
S.~Galal and M.~Horowitz, ``Energy-efficient floating-point unit design,''
  \emph{Computers, IEEE Transactions on}, vol.~60, no.~7, pp. 913--922, 2011.

\end{thebibliography}

%
\begin{IEEEbiography}[{\includegraphics[
  width=1in,height=1.25in,clip,keepaspectratio]{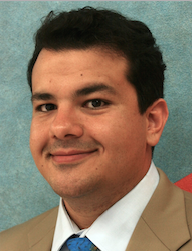}}
  ]{Ardavan Pedram}
 received his M.Sc. degree in computer engineering from University of Tehran in 2006 and his Ph.D. from The University of Texas at Austin in 2013. Currently, he is a Postdoctoral fellow at Stanford University. His research interests include high performance computing, and computer architecture. He specifically works on hardware-software co-design (algorithm for architecture) of special purposed accelerators for high-performance energy-efficient linear algebra, machine learning, and signal processing.
\end{IEEEbiography}

\begin{IEEEbiography}[{\includegraphics[
  width=1in,height=1.25in,clip,keepaspectratio]{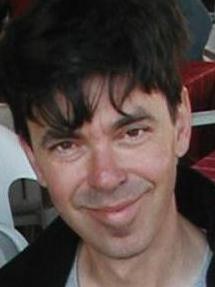}}
  ]{Stephen Richardson}
  holds a Ph.D. in Electrical Engineering from
  Stanford, and has worked in industry at Weitek and MIPS, as
  well as at Sun Microsystems and Hewlett-Packard research labs.
  Dr. Richardson is currently a Research Associate in the 
  Stanford University EE Department.
\end{IEEEbiography}

\begin{IEEEbiography}[{\includegraphics[
  width=1in,height=1.25in,clip,keepaspectratio]{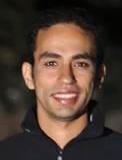}}
  ]{Sameh Galal} received the B.Sc. degree in Electronics Engineering and Computer Science from the American University in Cairo in 2005 and the Ph.D. degree in Electrical Engineering from Stanford University in 2012. His research
  interests include energy efficiency and floating point unit design. He currently works at Soft Machines Inc.
\end{IEEEbiography}

\begin{IEEEbiography}[{\includegraphics[
  width=1in,height=1.25in,clip,keepaspectratio]{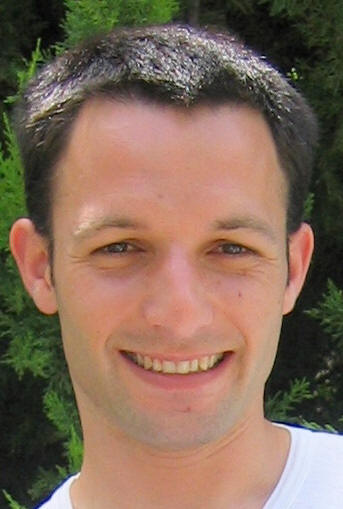}}
  ]{Shahar Kvatinsky}
is an assistant professor at the Electrical Engineering Department, Technion – Israel Institute of Technology. He received the B.Sc. degree in computer engineering and applied physics and an MBA degree in 2009 and 2010, respectively, both from the Hebrew University of Jerusalem, and the Ph.D. degree in electrical engineering from the Technion in 2014. From 2006 to 2009 he was with Intel as a circuit designer and was a post-doctoral research fellow at Stanford University from 2014 to 2015. Kvatinsky is an editor in Microelectronics Journal and has been the recipient of six Technion excellence teaching, and numerous other awards. His current research is focused on circuits and architectures with emerging memory technologies and design of energy efficient architectures.
\end{IEEEbiography}

\begin{IEEEbiography}[{\includegraphics[
  width=1in,height=1.25in,clip,keepaspectratio]{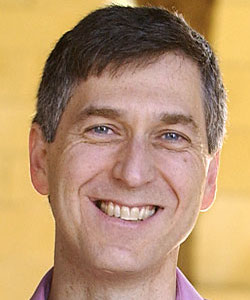}}
  ]{Mark Horowitz}
  is the Yahoo! Founders Professor at Stanford University
  and was chair of the Electrical Engineering Department from 2008 to
  2012. He co-founded Rambus, Inc. in 1990 and is a fellow of the IEEE
  and the ACM and a member of the National Academy of Engineering and
  the American Academy of Arts and Science. Dr. Horowitz's research
  interests are quite broad and span using EE and CS analysis methods to
  problems in molecular biology to creating new design methodologies for
  analog and digital VLSI circuits.
\end{IEEEbiography}







\end{document}